\documentstyle[prb,aps,multicol,epsfig]{revtex}
\begin{document}
\draft
\title{Electron transport in a quantum wire with realistic Coulomb
interaction}

\author{V. A.~Sablikov\cite{e-mail} and B. S.~Shchamkhalova}
\address{Institute of Radio Engineering and Electronics,
Russian Academy of Sciences, Fryazino, Moscow District,
141120, Russia}
\date{to appear in Phys. Rev. B}
\maketitle

\begin{abstract}
Electron transport in a quantum wire with leads is investigated with actual
Coulomb interaction taken into account. The latter includes both the direct
interaction of electrons with each other and their interaction via the image
charges induced in the leads. Exact analytical solution of the problem is
found with the use of the bosonization technique for one-dimensional
electrons and the three-dimensional Poisson equation for the electric field.
The Coulomb interaction is shown to change significantly the electron density
distribution along the wire as compared with the Luttinger-liquid model with
short-range interactions.  In dc and low-frequency regimes, the Coulomb
interaction causes the charge density to increase strongly in the vicinity of
the contacts with the leads. The quantum wire impedance shows an oscillating
behavior versus the frequency caused by the resonances of the charge waves.
The Coulomb interaction produces a frequency-dependent renormalization of
the charge wave velocity.

\vspace{0.5cm}
\noindent PACS number(s): 72.10.Bg
\end{abstract}

\vspace{1cm}
\multicols{2}
\section{Introduction}
The electron-electron interaction is generally recognized now to be
fundamentally important in one-dimensional (1D) structures. The interaction
of 1D electrons turns out to be so significant that the Fermi liquid concept
breaks down.  More adequate becomes the notion of a strongly correlated state
known as Luttinger liquid (LL) with bosonlike
excitations.~\cite{Haldane,Voit} The commonly used LL model treats the
electron-electron interaction as the short-range one. In this approach, the
electron-electron interaction changes the electron liquid compressibility
giving rise to a renormalization of charge-wave velocity.  Generally
speaking, the renormalization parameter $g$ is a function of the wave vector
$p$ of boson excitations. However, within the short-range interaction
approach the parameter $g$ is supposed to be a constant.  Presently
there is no unambiguous evidence that the LL really exists in quantum wires
or those rejecting this concept. The experiment of Tarucha {\it et
al.}~\cite{Tarucha} has shown that the dc conductance of a quantum wire
structure is quantized by standard steps $e^2/h$. This result was explained
in the frame of the LL model.~\cite{Maslov,Ponomarenko1,SafiSchulz} More
recently, Yacoby {\it et al.}~\cite{Yacoby1,Yacoby2} have found a
nonuniversal conductance quantization by steps different from $e^2/h$.
One can therefore conclude that the experiments on quantum wires reveal more
complex behavior of 1D conduction than the simple LL model predicts. Thus an
important problem is to develop the theory for actual quantum wire
structures. Application of the LL theory for this purpose points out two
problems.

First, the assumption that electrons interact with each other locally is
evidently inadequate in the real situation since the Coulomb interaction is
essentially nonlocal. This assumption is often
justified~\cite{GlazmanRuzinShklovskii,Egger} by the screening effect of the
highly conducting gate electrode. In this case the gate current should be
taken into account~\cite{Blanter} to provide the charge conservation and
cause the theory to be gauge invariant.~\cite{Buttiker} Hence the screening
effect of conducting electrodes (gates and leads) should be thoroughly
analyzed and taken into consideration in order to understand the
experimental situation. In reality, the screening by the electrodes consists
in the appearance of image charges of the electrons which are situated inside
the quantum wire.  Due to the image charges, the electron-electron interaction
becomes dipolelike (or generally multipolelike) but the dependence of $g$
on $p$ may be essential for the structures realized experimentally.

Second, the conductance of the mesoscopic quantum wire structure is
known~\cite{Maslov,Sandler} to be substantially determined by the contacts of
the wire with the leads. Besides, the contacts between the quantum wire and
the leads cause reflection of bosonlike excitations which determines the
high-frequency behavior of the
admittance.~\cite{SafiSchulz,Ponomarenko2,Sablikov} Thus, the interaction of
electrons moving in the quantum wire with leads has to be taken into
consideration.

The purpose of the paper is to obtain the actual form of the electron-electron
interaction potential in quantum wire structures with leads and to investigate the
phase-coherent transport of electrons in both the dc and ac regimes.
The difficulty of this problem is caused by a nonlocal nature of the
interaction and by the fact that in a quantum wire of finite length the
translational symmetry is broken and hence the electron-electron interaction
potential depends separately on the coordinates $x$ and $x'$ of interacting
electrons rather than on the difference $|x-x'|$. We have found a situation
in which this problem is solved exactly in the frame of the bosonization
technique.  It is realized when the lead surfaces may be approximated by
planes perpendicular to the wire.

It is worth noting that there is an alternative (contactless) approach in
studying the ac transport of electrons in quantum wires. In this case the
quantum wire is not supplied with leads. The ac transport is investigated by
means of measuring the absorption or scattering of the electromagnetic
radiation. This situation was recently considered by Cuniberti, Sassetti, and
Kramer~\cite{Cuniberti} for a homogeneous quantum wire of infinite length.
They have investigated ac conductance defined via the absorption of
electromagnetic radiation taking into account the electron-electron
interaction of finite range and an arbitrary distribution of the external
electric field along the wire. It was found that both the interaction length
and the electric field distribution affect significantly the ac conductance.
In the present paper we show that the leads produce an essential effect
due to inhomogeneity of the electron-electron interaction in the wire.
It manifests itself in the charge-density distribution along the wire
and the frequency dependence of the impedance.

The paper is organized as follows. In Sec. {\rm II} the potential of
electron-electron interaction in a quantum wire structure with leads is
obtained for the electron interaction with the charges induced on the leads
taken into account. Section {\rm III} describes the equation of motion for
bosonized phase field with nonlocal interaction and gives its solution via
expansion in terms of the eigenfunctions of the electron-electron interaction
potential. In Sec. {\rm IV}, charge-density distribution in the structure
is investigated for both the dc and ac regimes.  Section {\rm V} contains
the calculation and analysis of the impedance of the quantum wire structure.

\section{Electron-electron interaction potential}
The mesoscopic structure under consideration consists of a quantum wire
coupled to two bulky (2D or 3D) regions (the electron reservoirs) which serve
as leads.  The electrons in the wire interact with each other both directly
and via the surface charges which are induced on the surface of the leads.
The electron-electron interaction energy $W$ is defined by the product of the
electron density $\rho({\bf r})$ at a point ${\bf r}$ and the potential
$\varphi_i(\bf r,\bf r')$ created at this point by the charge at a point
${\bf r'}$.  This potential is determined by the Laplace equation with
boundary conditions corresponding to the given configuration of the leads.
When calculating $\varphi_i(\bf r,\bf r')$, it is reasonable to consider the
lead surfaces as equipotential ones. This is a natural assumption. As we are
interested mainly in the electron behavior in the quantum wire, we can assume
that the characteristic times of electron processes (such as Maxwell
relaxation and plasma waves) inside the reservoirs are much shorter than the
electron transit time through the quantum wire. This will be the case if the
reservoirs are perfectly conducting.

Distribution of the electron density $\rho({\bf r})$ in the channel can be
written in the form
\begin{equation} \label{rho-rx}
\rho({\bf r}) = \chi({\bf r}_{\perp}) \rho(x)\:,
\end{equation}
where $\chi({\bf r}_{\perp})$ is a normalized function of the radial
coordinate perpendicular to the channel and $\rho(x)$ is a function of the
coordinate along the channel.

The potential $\varphi_i({\bf r,\bf r'})$ created by an electron density is
expressed via the Green function of the Laplace equation with zero
boundary conditions at the lead surfaces.

The product of $\rho({\bf r})$ and $\varphi_i({\bf r,\bf r'})$ can be
integrated over the transverse coordinates ${\bf r}_{\perp}$ and  ${\bf
r'}_{\perp}$ to give the following expression for the electron-electron
interaction energy:~\cite{SablShch}
\begin{equation} \label{W2}
W = \frac{1}{2}\int\!\int\,dx\,dx'\, \rho(x)\, \rho(x')\,U(x,x').
\end{equation}

\begin{figure}[htb]
\mbox{\epsfig{file=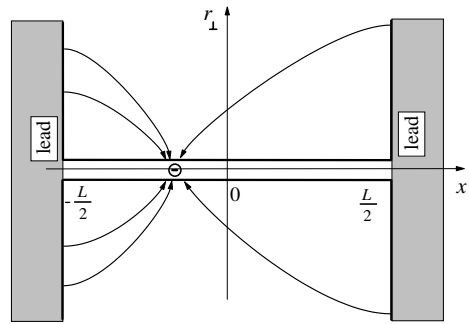,width=8cm}}
\narrowtext
\caption[to]{Schematic view of a quantum wire structure with leads,
a moving charge, and electric force lines.}
\label{fig1}
\end{figure}
An explicit form of $U(x,x')$ was found in Ref.~\onlinecite{SablShch} for
realistic situation where the electrode surfaces are two planes $x=-L/2$ and
$x=L/2$ perpendicular to the channel (Fig.~1). In this case
\begin{equation} \label{U(x,x')}
U(x,x') = \frac{e^2}{\pi \epsilon}\,\int d^2q\,|\chi_q|^2\,G_q(x,x'),
\end{equation}
$\epsilon$ is the dielectric constant outside of the channel, $\chi_q$ is the
Fourier-Bessel transform of the radial function $\chi({\bf r}_{\perp})$,
and
\endmulticols
\widetext
\vspace{-6mm}\noindent\underline{\hspace{87mm}}
\begin{equation} \label{G_q(x,x')}
G_q(x,x') = \frac{L}{qL{\rm sinh}(qL)} \left\{
\begin{array}{rcl}
{\rm sinh}\left[q\left(L/2 + x\right)\right]\,{\rm sinh}\left[q\left(L/2 -
x'\right)\right] & {\rm if}& x<x',\\
{\rm sinh}\left[q\left(L/2 - x\right)\right]\,{\rm sinh}\left[q\left(L/2 +
x' \right)\right]\, & {\rm if}& x>x'.\\
\end{array}
\right.
\end{equation}
\noindent\hspace{92mm}\underline{\hspace{87mm}}\vspace{-3mm}
\multicols{2} \noindent
The interaction potential defined by Eqs.~(\ref{U(x,x')}) and
(\ref{G_q(x,x')}) is shown in Fig.~2 as a function of $x$ for a variety of
$x'$, with $V(x,x')$ being a normalized form of $U(x,x')$:
$$
V(x,x') = U(x,x')\,\epsilon L/e^2\,.
$$
\begin{figure}[htb]
\narrowtext
\hspace{-1cm}\mbox{\epsfig{file=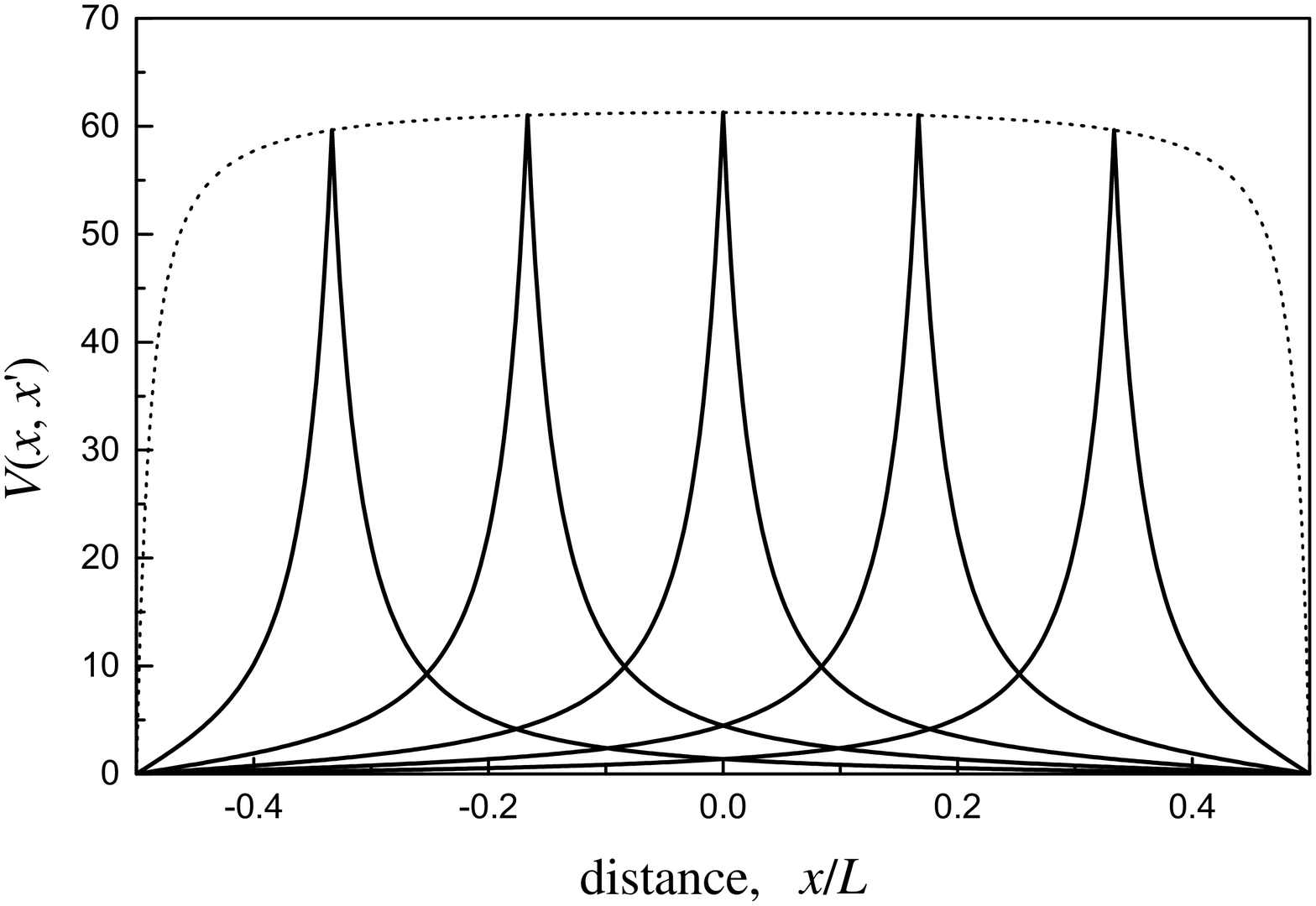,width=9.5cm}}
\caption[to]{Distance dependence of the electron-electron interaction
potential for a variety of $x'$. Dotted line is the envelope of $V(x,x)$
maxima.}
\label{fig2}
\end{figure}

When the interelectron distance is larger than the wire radius $a$
($|x-x'|\gg a$), the potential decreases as $V\sim L/|x-x'|$. In the middle
part of the quantum wire $V(x=x')\sim L/a$. Near the contacts
($x,x'\to \pm L/2$), $V$ goes to zero due to screening effect of the charges
induced on the lead surfaces. The behavior of this kind is quite general
for the interaction potential regardless of the specific configuration of the
leads. The potential defined by Eqs.~(\ref{U(x,x')}) and (\ref{G_q(x,x')})
will be used below in getting an exactly solvable model of the interacting
electron transport.

\section{The equation of motion}
To study a linear response of the quantum wire structure to an external
voltage $V_a \exp (-i\omega t)$ applied across the leads, we can restrict
ourselves to the consideration of low-energy excitations of the electron
system.  The most adequate method for this purpose is the standard
bosonization technique.~\cite{Haldane,Voit} We will use this technique
assuming that the electron density fluctuations are long-range ones.

When external voltage is applied across the electrodes, the electric
potential $\varphi(\bf r)$ in the wire is determined by the Poisson
equation with the boundary conditions controlled by the applied voltage:
$\varphi = 0$ at the left reservoir and $\varphi = V_a$ at the right
reservoir. It is convenient to present the electric potential in the wire as
a sum
\begin{equation}
\label{potential}
\varphi = \varphi_{\rm ext} + \varphi_{i}\,.
\end{equation}
Here $\varphi_{\rm ext}$ is the potential which would be without the wire. It is
determined by the Laplace equation with the same boundary conditions as the
total potential $\varphi$. The potential $\varphi_{i}$ is defined by the
Poisson equation with zero boundary conditions. The potential $\varphi_{i}$
is precisely the one used in Sec.~{\rm II} while calculating
the electron-electron interaction energy [see Eqs.~(\ref{W2}) and
(\ref{U(x,x')})].

Thus the bosonized Hamiltonian~\cite{Voit,Schulz} of 1D spinless electrons
can be taken in the form
\endmulticols
\widetext  
\vspace{-6mm}\noindent\underline{\hspace{87mm}}
\begin{equation}
H=\int\!\frac{dx}{2\pi}\,v_F\!\left[(1\!+\!g_1)\pi^2\Pi^2 +
(1\!-\!g_1)\left(\partial_x \Phi\right)^2\right]
-e\int dx\,\rho(x,t)\varphi_{\rm ext}(x,t) +
\!\int\!\!\!\int\!\frac{dx\,dx'}{2\pi^2}\,(\partial_x\Phi)\,U(x,x')\,
(\partial_{x'}\Phi)\,,
\label{hamiltonian}
\end{equation}
where $\Phi(x,t)$ is the phase field related to the charge excitations,
$\Pi(x,t)$ is the momentum density conjugate to $\Phi$, $v_F$ is the Fermi
velocity, and $g_1$ is the backscattering  parameter.~\cite{Schulz}

By writing this Hamiltonian, we assume implicitly that the ground state is
uniform. More careful investigation \cite{SabPol1} shows that really
the ground state is nonuniform due to two factors: ({\em i}) charging of the
quantum wire which occurs as a consequence of the electron transfer between
the wire and the reservoirs during the process of the establishment of the
equilibrium electrochemical potential and ({\em ii}) Friedel oscillations
near the contacts. The charge stored in the wire depends on the wire radius
and the background density of the positive charge. This charge may be both
positive and negative.  Under certain conditions the wire remains neutral.
The charging effect will be investigated elsewhere. In the present paper the
ground state is assumed to be neutral.  The Friedel oscillations of the
electron density have a characteristic length of the order of the Fermi
wavelength. Since we restrict our consideration by the long-range variation
of the electron density, the ground state may be considered as uniform.

The long wave component of the electron density $\rho$ is related to
$\Phi$ by
$$
\rho = -\frac{1}{\pi} \: \partial_x \Phi\;.
$$

From the Hamiltonian (\ref{hamiltonian}) we get the following equation of
motion for the phase field $\Phi$:
\begin{equation} \label{Phi(x,t)}
\partial_t\left(\frac{1}{v \tilde g}\partial_t \Phi \right) -
\partial_x\left(\frac{v}{\tilde g}\partial_x \Phi \right) =
\partial_x\left[e\:\varphi_{\rm ext} + \frac{1}{\pi}\,
\int\limits_{-L/2}^{L/2}dx'\,U(x,x')\,\partial_{x'} \Phi\right]\;,
\end{equation}
\noindent\hspace{92mm}\underline{\hspace{87mm}}\vspace{-3mm}
\multicols{2} \noindent
with
$$
v = v_F \sqrt{1-g_1^2}\;,\qquad\qquad
\tilde g = \sqrt{\frac{1+g_1}{1-g_1}}\;.
$$

Following Refs. 4-6,  we extend the
one-dimensional Eq.~(\ref{Phi(x,t)}) to the reservoirs assuming that the
electron density inside them is extremely high and their conductivity is
ideal. In such a case, $\varphi_{\rm ext}$ is a constant inside the reservoirs.
Hence the first term on the right-hand side of Eq.~(\ref{Phi(x,t)}) can be
omitted.  The second term (appearing from the electron-electron interaction
energy $W$) is known to be small as compared with the kinetic energy on the
left-hand side of Eq.~(\ref{Phi(x,t)}) when the electron density is high. In
this sense, the electrons in the reservoirs are noninteracting although the
external field is of course ideally screened there. Thus, the right-hand side
of Eq.~(\ref{Phi(x,t)}) may be dropped in the reservoirs. The solution of the
bosonized equation in the reservoirs should satisfy the condition that the
density wave be restricted at $|x|\to \infty$ when the external voltage is
turned on adiabatically.  The boundary conditions at the contacts between the
wire and the leads require the continuity of $\Phi$ and the particle current.
The latter means that $(v/\tilde g)\partial_x \Phi$ must be continuous there.

We find the exact solution of Eq.~(\ref{Phi(x,t)}) with the electron-electron
interaction potential $U(x,x')$ defined by Eqs.~(\ref{U(x,x')}) and
(\ref{G_q(x,x')}). In terms of dimensionless variables
$$
\xi=\frac{x}{L}\:,\qquad u=\frac{v\,\Phi}{e\tilde gLV_a}\:,\qquad
f=\frac{\varphi_{\rm ext}}{V_a}\:,
$$
Eq.~(\ref{Phi(x,t)}) takes the form:
\begin{equation} \label{u(xi)}
\frac{d}{d \xi}\left[\frac{du}{d \xi} + \beta \hat V \frac{du}{d \xi}
- f(\xi)\right]\, + \,\Omega^2 u\, =\, 0\:.
\end{equation}
Here
$$
\beta = \frac{e^2 \tilde g}{\pi\,\epsilon\,v}\:,\qquad \qquad
\Omega = \frac{\omega L}{v}\:,
$$
the operator $\hat V$ is defined as
\begin{equation} \label{V}
\hat V \psi = \int\limits_{-1/2}^{1/2} d\xi'\:V(\xi ,\xi ')\:\psi(\xi ')\:,
\end{equation}
where
\endmulticols
\widetext
\vspace{-6mm}\noindent\underline{\hspace{87mm}}
\begin{equation}\label{vv}
V(\xi ,\xi ') = 2\,\int\limits_0^{\infty}
\frac{dy}{{\rm sinh}y}\,|\chi_y|^2\:\left \{
\begin{array}{rcl}
{\rm sinh}\left[y\left(1/2 + \xi \right)\right]\,{\rm sinh}\left[y\left(1/2 -
\xi '\right)\right] & {\rm if}& \xi <\xi ',\\
{\rm sinh}\left[y\left(1/2 - \xi \right)\right]\,{\rm sinh}\left[y\left(1/2 +
\xi ' \right)\right]\, & {\rm if}& \xi >\xi '.\\
\end{array}
\right.
\end{equation}
\noindent\hspace{92mm}\underline{\hspace{87mm}}\vspace{-3mm}
\multicols{2} \noindent

The electron density in standard units is related to $u_{\xi}$ by
\begin{equation} \label{rho(x)}
\rho (\xi) = - \frac{2\,e\,\tilde g\,V_a}{h\,v}\,u_{\xi}(\xi)\;.
\end{equation}

After solving Eq.~(\ref{u(xi)}) in the reservoirs we use the continuity
of the electron flow and the phase $\Phi$ at the contacts to obtain the
boundary conditions directly to Eq.~(\ref{u(xi)}) in the inner region of
the quantum wire, $-1/2 \le \xi \le 1/2$,
\begin{equation}
\label{boundary}
u_{\xi}\,\pm \,i\,\Omega \tilde g \,\left.u\right| _{\xi =\mp 1/2}\,=\,0\;.
\end{equation}

The integro-differential equation of the form Eq.~(\ref{u(xi)}) may be solved
via an expansion in terms of the eigenfunctions of the operator $\hat V$.
It is easy to verify that the functions
$$
\psi_n(\xi) = \sqrt 2 \sin \left[\pi n\left(\xi+\frac{1}{2}\right)\right] \:,
\qquad n=1,2,3,\dots \:,
$$
are the eigenfunctions of $\hat V$  defined by Eqs.~(\ref{V}) and (\ref{vv})
with the eigenvalues
\begin{equation} \label{lambda_n}
\lambda_n =  2 \int_0^{\infty} \frac{y\,dy}{y^2+(n\pi )^2}\; |\chi_y|^2 \;.
\end{equation}

For the geometry of the sample under consideration, the external potential
$f(\xi)$ is a linear function which may be expanded in terms of $\psi_n(\xi)$
with even $n$.

The exact solution of Eq.~(\ref{u(xi)}) with the boundary
conditions~(\ref{boundary}) can be obtained via the expansion in terms of
$\psi_n(\xi)$. As a result, we get the following expressions for the
dimensionless electron density $u_{\xi}(\xi)$:
\begin{equation}
\label{u'M}
u_\xi(\xi)\,=\,\sum_{n=1}^{\infty} c_n \sin\left[2\pi n(\xi+1/2)\right]\,
\end{equation}
and the phase field $u(\xi)$,
\begin{equation}
\label{uM}
u(\xi)\,=\,A(\Omega) - \sum_{n=1}^{\infty} \frac{c_n}{2\pi n}
\cos\left[2\pi n(\xi+1/2)\right]\,,
\end{equation}
where
\begin{equation} \label{c_n}
c_n = B(\Omega)\: \frac{4\pi n}{4\pi^2n^2(1+\beta \lambda_{2n})-\Omega^2}\,,
\end{equation}
\begin{equation}
A(\Omega) = \frac{1}{\Omega}\:\frac{1-4i\tilde g\Omega D(\Omega)}
{2i\tilde g + \Omega - 4i\tilde g\Omega^2 D(\Omega)}\,,
\label{A}
\end{equation}
\begin{equation}
B(\Omega)  = \frac{-2i\tilde g}{2i\tilde g + \Omega - 4i\tilde g\Omega^2
D(\Omega)}\:,
\label{B}
\end{equation}
\begin{equation}
D(\Omega) = \sum_{n=1}^{\infty}\frac{1}{4\pi^2n^2(1+\beta
\lambda_{2n}) - \Omega^2}\,.
\label{D}
\end{equation}

\section{Electron density distribution}
The purpose of this section is to clarify how the real Coulomb interaction
affects the value and the distribution of the charge density in the
quantum wire structure.

First of all, let us consider a limiting case of the short-range interaction.
In this case, $V(x,x')\propto \delta(x-x')$ and hence the eigenvalues are
independent of $n$ and $\lambda_{2n}=\lambda$. All the sums are easily
calculated, which results in the following expression for the normalized
electron density:
\begin{equation}\label{u_LL}
u_{\xi}(\xi)=\frac{1}{\Omega\sqrt{1+\beta\lambda}}\:
\frac{g^*\sin(\Omega^* \xi)}
{g^*\cos(\Omega^*/2)-i\sin(\Omega^*/2)},
\end{equation}
where renormalized values $g^*=\tilde g/\sqrt{1+\beta\lambda}$ and
$\Omega^*=\Omega/\sqrt{1+\beta\lambda}$ are introduced.
The density $\rho(\xi)$ calculated according to Eq.~(\ref{u_LL}) coincides
exactly with that found in Ref.~\onlinecite{Sablikov} in the framework of
the standard LL model with the interaction parameter $g=g^*$.

In the limit of $\Omega \to 0$, Eq.~(\ref{u_LL}) yields
\begin{equation}\label{u_LL0}
u_{\xi}(\xi)=\frac{\xi}{1+\beta\lambda}\:.
\end{equation}
With increasing, frequency the charge waves appear which have
resonances~\cite{Sablikov} along the wire when $\Omega=2\pi n
\sqrt{1+\beta\lambda}$.

Another case will be useful in what follows as a reference point to
demonstrate the Coulomb interaction effect. It is the case of
noninteracting electrons which corresponds to $\beta =0$ in
Eqs.~(\ref{u_LL}) and (\ref{u_LL0}).

For the case of the realistic interaction, the electron density distribution
given by Eq.~(\ref{u'M}) is generally more complicated. However, simple
results are obtained for the regions near the contacts and in the middle
part of the wire, taking into account the specific behavior of $\lambda_n$
versus $n$.  It is determined by the fact that the radial function
$\chi({\bf r}_{\perp})$ is located in the region of radius $a$ which is much
shorter than the wire length $L$, i.e., $\alpha = a/L$  is a small parameter.
The results which will be given below are qualitatively valid for any
localized function $\chi({\bf r}_{\perp})$. To be specific, we will
use the Gaussian form for $\chi({\bf r}_{\perp})$ when it is necessary to
bring the calculations to final form. One obtains that $\lambda_n$ varies
slowly with $n$ for $\pi \alpha n\ll 1$ and $\lambda_n$ decreases as
$n^{-2}$ for $\pi \alpha n\gg 1$.

In the vicinity of the contacts, the main contribution to the sum in
Eq.~(\ref{u'M}) is due to the large-$n$ terms for which an asymptotic
expression of $\lambda_n \sim n^{-2}$ can be used. Thus the
following expression for the normalized electron density is obtained in the
vicinity of the left electrode [$(1/2 + \xi)\ll 1$]:
\begin{equation} \label{uu'1/2} 
u_{\xi}(\xi)\approx - \frac{B(\Omega)}{2}
\frac{\sinh(\xi\sqrt{2\beta/\alpha^2-\Omega^2})}
{\sinh(\sqrt{2\beta/\alpha^2-\Omega^2}/2)} \,,
\end{equation} 
with $B(\Omega)$ being defined by Eq.~(\ref{B}). Equation~(\ref{uu'1/2}) shows
that
({\it i}) $u_{\xi}$ decreases with the distance from the contact as
$\sinh(x/\ell)$, with characteristic length being
$$
\ell\,=\,\frac{a}{\sqrt{2\beta-\alpha^2\Omega^2}}\;,
$$
({\it ii}) at $\Omega=0$, the boundary value of $u_{\xi}$ is
equal to 1/2 and is independent of the interaction.~\cite{SablShch}

In the middle part of the wire, Eq.~(\ref{u'M}) may be simplified when
$\alpha$ is exponentially small, i.e., $-\ln \alpha \gg 1$. In this case the
sum in Eq.~(\ref{u'M}) may by estimated assuming
$$
\lambda_n \approx \lambda_0\equiv -\ln(2\pi^2\alpha^2) - \gamma\,,
$$
$\gamma \approx 0.577 21\dots$ being Euler's constant.
This calculation results in the same equation as Eq.~(\ref{u_LL})
for the short-range interacting electron gas, where $\lambda$ should be
replaced by $\lambda_0$.  This is equivalent to introducing an effective
interaction parameter
$$
g_{\rm eff}\approx\, \frac{\tilde g}{\sqrt{1+\beta \lambda_0}}
$$ into the LL model.

For the sake of simplicity, we suppose $\tilde g=1$ below.

\begin{figure}[htb]
\narrowtext
\hspace{-1cm}\mbox{\epsfig{file=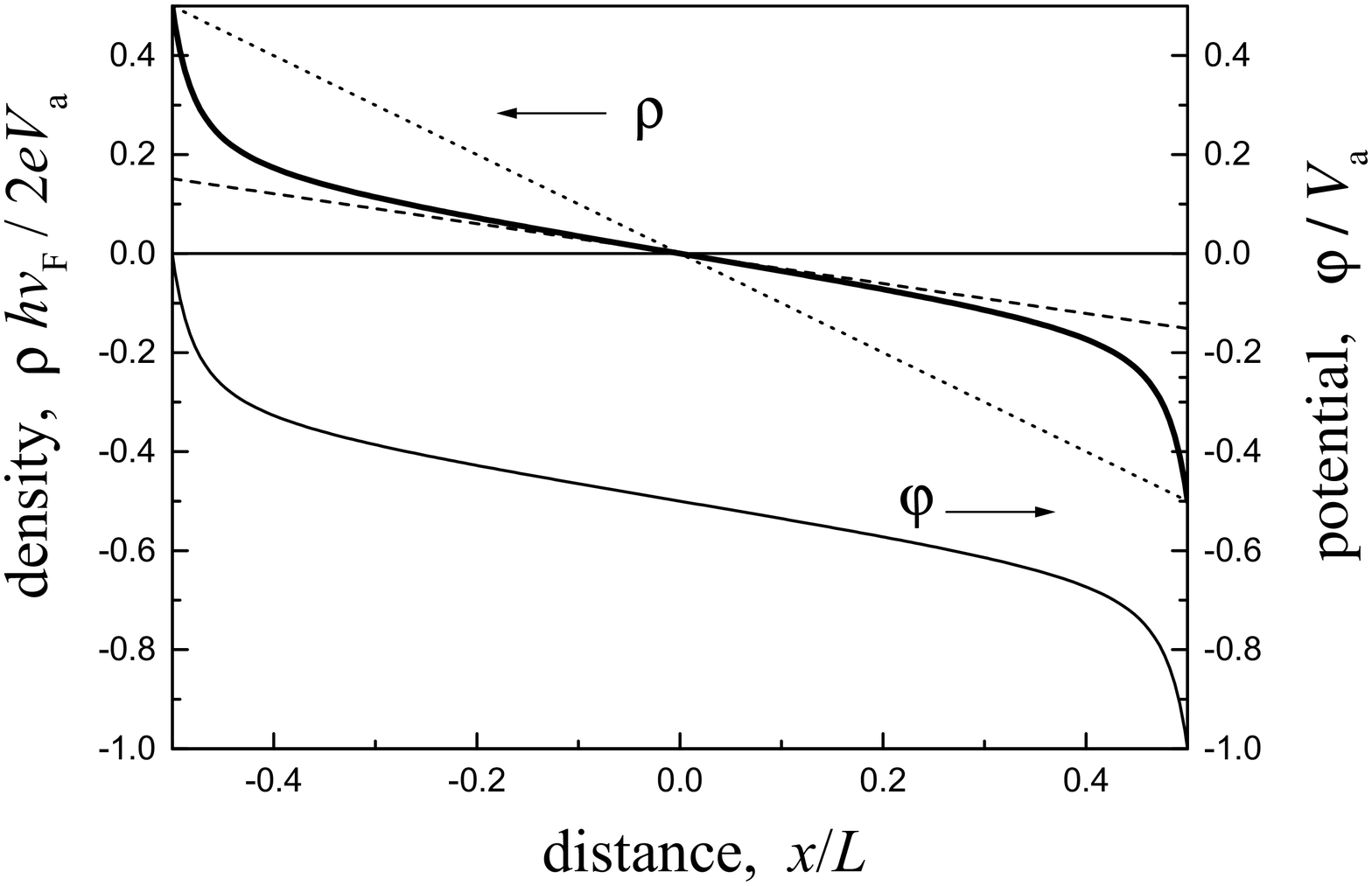,width=9.5cm}}
\caption[to]{Distribution of the dc electron density $\rho$ (thick solid
line) and the potential $\varphi$ (thin solid line) along the wire for
long-range Coulomb interaction.  The dashed line is the density
distribution according to the LL model with $g_{\rm eff}$, the dotted line is
that for noninteracting electrons.  The parameters used are: the Fermi
energy $\varepsilon_F =5$~meV and $\alpha =0.02$.  The interaction parameter
$\beta$ was calculated to be 0.35.}
\label{Fig.3}
\end{figure}

The effect of the Coulomb interaction on the electron density distribution in
the quantum wire is demonstrated by Fig.~3 for the dc condition when a
positive potential is applied to the left electrode with respect to the right
one.  Under the action of the external electric field, the electron system in
the quantum wire is polarized: the electron liquid is compressed in the left
part of the wire and decompressed in the right one. Respectively, at the left
end of the quantum wire the excessive electrons appear while at the right end
the electron density is decreased.  If the electron-electron interaction is
omitted [see the dotted line in Fig.~3 and Eq.~(\ref{u_LL0}) with $\beta =0$],
the electron density decreases linearly with the distance, with the boundary
value of the normalized density being equal to $\pm 1/2$. When the
short-range interaction with the effective interaction parameter $g_{\rm
eff}$ is turned on, the density distribution remains linear (dashed line).
But the slope goes down as a result of a decrease in the electron liquid
compressibility.  Note that the boundary value of the electron density also
decreases.

When the actual Coulomb interaction is turned on, the electron density
distribution is changed qualitatively (thick line in Fig.~3). As compared
with the noninteracting case, the charge density in the middle part of the
wire decreases, which may be interpreted as neutralization of the negative
and positive charges due to their mutual attraction. Towards the contacts the
charge density increases reaching $\pm 1/2$ at the boundaries. This behavior
may be understood from the fact that near the contacts the charges in the
wire are neutralized by the image charges in the electrodes.

On the other hand, if we compare the Coulomb interaction case with the
short-range interaction model, we find that the actual Coulomb interaction
leads to an increase of the electron density fluctuation near the contacts.
This fact may be interpreted as a result of the decrease in the interaction
parameter $g$ due to screening the electron-electron interaction by the
electrode.

At finite frequency, this near-contact effect of the Coulomb interaction
is preserved up to a characteristic frequency
$\Omega_w=\sqrt{2\beta}\,/\alpha$, above which the exponentially decreasing
part of $\rho(x)$ disappears.

The electric potential $\varphi$ is easily calculated using
Eq.~(\ref{potential}), Green function (\ref{G_q(x,x')}), and the electron
density $\rho(\bf r)$ found above. Distribution of $\varphi (\xi)$ along the
quantum wire is shown in Fig.~3. A good proportionality is found between
$\rho$ and $\varphi$ for dc conditions when $a/L \ll 1$:
$$
\rho(\xi) \approx {\rm const} + \frac{\tilde g}{\pi v} e\varphi(\xi)\,.
$$

The main effect observed when the frequency grows is the appearance
of the traveling charge waves. This is illustrated by Fig.~4, where the real
part of the normalized electron density is shown as a function of the
distance for a number of frequencies. The curves shown in Fig.~4 were
obtained by numerical calculation of Eq.~(\ref{u(xi)}).~\cite{SablPol} The
frequency is given in a normalized form
$$
\nu = \frac{\omega L}{2\pi v_F}
$$
as labels to the curves.

\begin{figure}[htb]
\narrowtext
\hspace{-1cm}\mbox{\epsfig{file=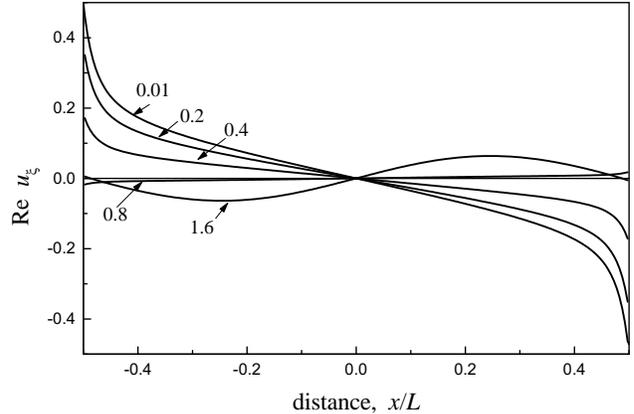,width=9.5cm}}
\caption[to]{Electron density distribution along the wire for a set of
frequencies. The curves are labeled by the normalized frequency $\nu$.
The parameters used in the calculations are the same as in Fig.~3.}
\label{fig4}
\end{figure}

Our analytical solution given by Eqs.~(\ref{u'M}) and (\ref{c_n}) shows that
there is a set of characteristic frequencies $\Omega_n$ which are determined
by the poles of $D(\Omega)$:
\begin{equation}
\Omega_n = 2\pi n \sqrt{1+\beta \lambda_{2n}}\,.
\label{Omega_n}
\end{equation}
For $\Omega = \Omega_n$, the electron density is a standing wave with
zeros at the contacts, $n$ being the wave number:
$$
\left(\frac{du}{d\xi}\right)_n \propto
\frac{\sin\left[2\pi n(\xi +1/2)\right]}{4\pi n (1+\beta \lambda_{2n})}\,.
$$
It is noteworthy that the electron flow turns to zero at the contacts
simultaneously with the electron density. Thus $\Omega_n$ is the resonant
frequencies of the charge waves in the quantum wire. Under the resonant
condition the electron density perturbation is locked inside the wire.

Equation~(\ref{Omega_n}) shows that the frequency $\Omega_n$ depends on
the wave number $2\pi n/L$ in a nonlinear manner due to the dependence of
$\lambda_n$ upon $n$. For $n\to \infty$, the resonant frequency is
proportional to the wave number which corresponds to the soundlike
dispersion. For low wave numbers [$n\le(\pi \alpha)^{-1}$], $\Omega_n$ is
noticeably higher than one expects from the soundlike dispersion.

It is instructive to compare Eq.~(\ref{Omega_n}) with the dispersion
equation for charge waves in an infinite quantum wire which was found by
Schulz~\cite{Schulz} using the bosonization technique,
\begin{equation}\label{Sch}
\Omega(p) = p\sqrt{1+\beta V_p}\,,
\end{equation}
where $p$ is the wave vector normalized by $L^{-1}$ and $V_p$ is the Fourier
transform of the interaction potential, which is approximated by the
modified Bessel function $K_0(\alpha p)$. A similar dispersion law was
obtained by Das Sarma and Hwang~\cite{DasSarma} for 1D plasmons in the
long-wavelength limit in the frame of the random-phase approximation taking
into consideration the more general form of the interaction potential.
Equation~(\ref{Sch}) differs
from Eq.~(\ref{Omega_n}) in the replacement $\lambda_{2n}$ by $V_p$.
A reasonable approximation for $\lambda_{2n}$ is
\begin{equation}
\label{bda}
\lambda_{2n} \approx \exp \left(2\pi^2 \alpha^2 n^2\right)\,
E_1\left(2\pi^2 \alpha^2 n^2\right)\,,
\end{equation}
where $E_1(z)$ is the exponential integral. It is worth noting that in the
limit $L\to \infty$, Eq.~(\ref{bda}) is the same as the Fourier transform of
the interaction potential used in Ref.~\onlinecite{Cuniberti} when the
screening length is much larger than the quantum wire diameter.  On setting
$p=2\pi n$ and comparing the expressions (\ref{Omega_n}) and (\ref{Sch}), one
can see that they are close when $\pi \alpha n \ll 1$ and differ
significantly in the opposite case.

One can say that Eq.~(\ref{Omega_n}) is a discrete version of the dispersion
relation for 1D electrons which takes correctly into account the
electron-electron interaction in a finite 1D system. It will be shown below
that the discrete character of the resonant frequencies $\Omega_n$ of finite
quantum wire results in strong peculiarities of the frequency dependence of
admittance.

\section{The impedance}
The electron current in a quantum wire is determined by the time derivative of
the phase field.~\cite{Voit} In the terms of the normalized phase $u$, the
current is
\begin{equation}
j(x,\omega) = -\frac{2i\omega e^2\tilde gL}{hv}\,u(x,\omega)\,V_a\,.
\label{j(x)}
\end{equation}
The current calculated in such a way depends on the coordinate $x$ along the
wire. However, the electric current $j_{\rm meas}$ which is detected by a
measuring device is obviously independent of $x$. This current is defined as
a charge flow through the leads. Its formation process is analyzed in
Appendix A as applied to the specific quantum wire structure considered here.
The measured current is the sum of a trivial capacitance current and the
current caused by the quantum wire presence.  According to the Shockley
theorem,~\cite{Shockley} the latter is
$$
j_{\omega} = \frac{1}{L}\int_{-L/2}^{L/2}\!dx\,j(x,\omega)\,.
$$

Due to Eqs.~(\ref{uM}) and (\ref{j(x)}), the current $j_{\omega}$ becomes
$$
j_{\omega} = i\frac{e^2}{h}\,2\tilde g\,A(\Omega)\,V_a\,,
$$
with $A(\Omega)$ being defined by Eq.~(\ref{A}).
This results in the following expression for the quantum wire structure
impedance:
\begin{equation}
\label{Z}
Z(\Omega)=\frac{h}{e^2}\left[\frac{1}{1 - 4i\tilde g\Omega D(\Omega)}
-\frac{i\Omega}{2\tilde g}\right]\,.
\end{equation}

In what follows, the impedance is analyzed rather than the admittance which is
usually considered, because the frequency dependence of the impedance shows
more pronounced features caused by the charge waves. The real part of $Z$ is
\begin{equation}
{\rm Re}Z = \frac{h}{e^2}\:\frac{1}{1+[4\tilde g\Omega D(\Omega)]^2} \,.
\label{ReZ}
\end{equation}
When $\Omega \to 0$, the impedance is equal to $h/e^2$  and is independent of
the interaction parameter $\beta$.  The frequency dependence of ${\rm Re}Z$
is mainly governed by that of $D(\Omega)$. The resonant frequencies
$\Omega_n$ are by definition the poles of $D(\Omega)$.  Between the
neighboring poles, there is a zero of $D(\Omega)$.  Equation~(\ref{ReZ}) shows
that ${\rm Re}Z=0$ when $|D|\to \infty$ and ${\rm Re}Z=h/e^2$ when $D\to 0$.
Thus with increasing frequency, ${\rm Re}Z$ oscillates  between zero (which
occurs at the resonant frequencies) and $h/e^2$, these limiting values being
independent of the interaction. The frequency dependence of ${\rm Re}Z$ is
illustrated by Fig.~5, where three cases are compared: noninteracting
electrons, the LL model with short-range interaction, and the electrons with
actual Coulomb interaction.

\begin{figure}[htb]
\narrowtext
\hspace{-1cm}\mbox{\epsfig{file=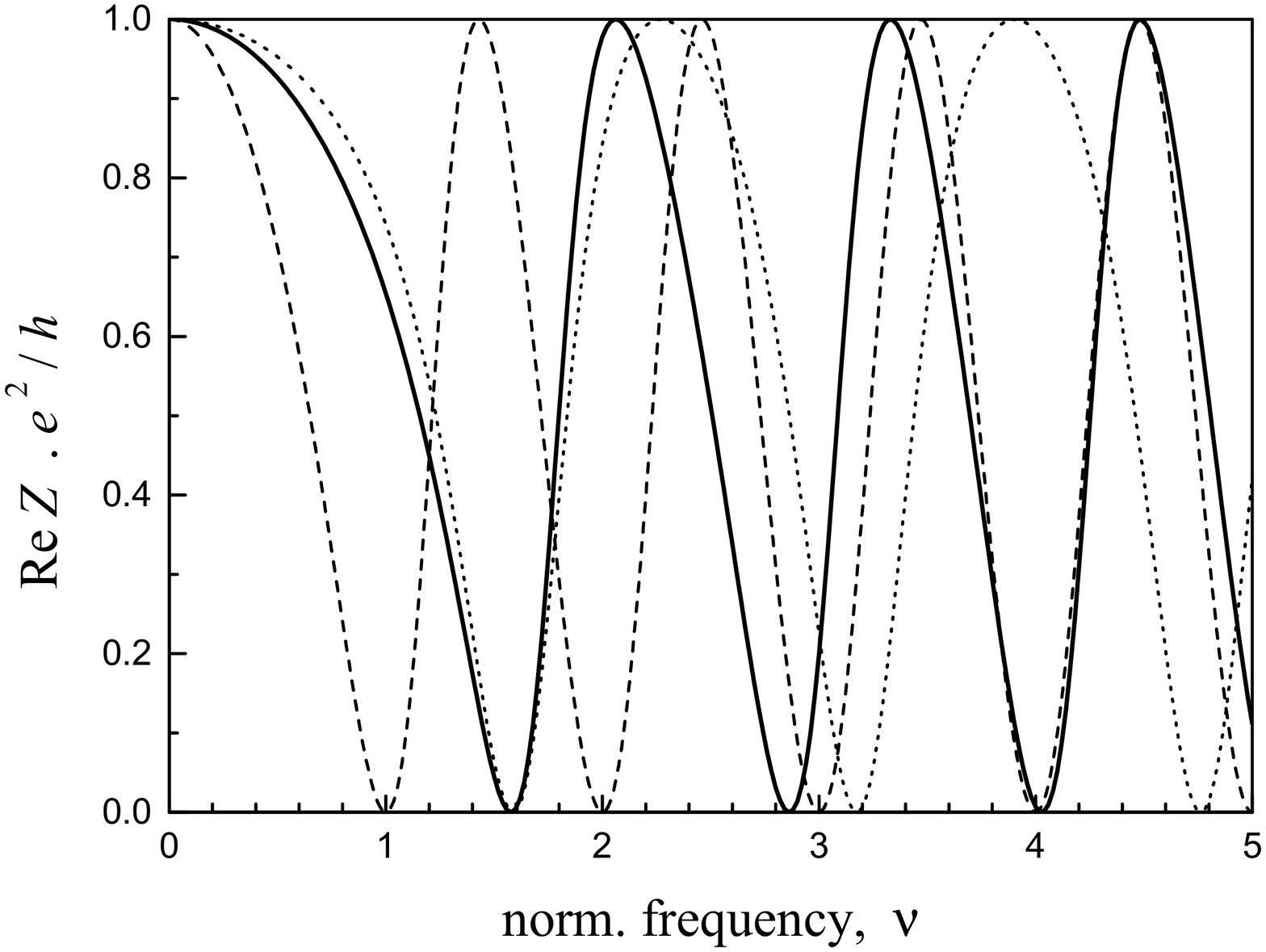,width=9.5cm}}
\caption[to]{Real part of the impedance versus the normalized
frequency $\nu$ for true long-range interaction (---), LL
approach with $g$=0.63 (- - -), and noninteracting electrons ($\cdots$).}
\label{fig5}
\end{figure}

An interesting result is that ${\rm Re}Z=0$ for the resonant frequencies.
Under resonant condition the time-dependent evolution of the electron density
is essentially oscillations between the two ends of the wire, and electrons
are not emitted/absorbed by the contacts. As a consequence, the electric
current component in-phase with the applied voltage vanishes, and the real
part of both the admittance and the impedance turns to zero. However, in
reality there is a finite dissipation which was not taken into account.  The
inclusion of the dissipation into calculations should restrict the minimum
of ${\rm Re}Z$ by some low value.

The resonances of ${\rm Re Z}$ occur when the frequency is a multiple of the
inverse time of flight of electron excitations along the quantum wire.
This conclusion was confirmed for both the
noninteracting~\cite{Velicky,Sablikov3} electrons and the short-range
interacting electrons in the LL model.~\cite{Sablikov} The fact that in the
case of the Coulomb interaction the impedance oscillations are nonperiodic
may be interpreted as a result of the frequency-dependent renormalization of
the charge-wave velocity due to the Coulomb interaction.  At low frequency,
the velocity renormalized by the Coulomb interaction is essentially larger
than $v_F$.  The resonance frequency spectrum shows that the velocity
decreases with frequency. It is worthwhile to note that the phase velocity is
important since the resonant conditions are obviously related to the wave
interference.

The imaginary part of the impedance may be presented in the form
\begin{equation}
\label{ImZ}
{\rm Im}Z = -\omega L_{\rm eff} + \frac{h}{e^2}\:
\frac{4\tilde g\Omega D(\Omega)}{1+[4\tilde g\Omega D(\Omega)]^2}\,,
\end{equation}
where
$$
L_{\rm eff}=\,\frac{hL}{2e^2v_F}\,.
$$
\begin{figure}[htb]
\hspace{-1cm}\mbox{\epsfig{file=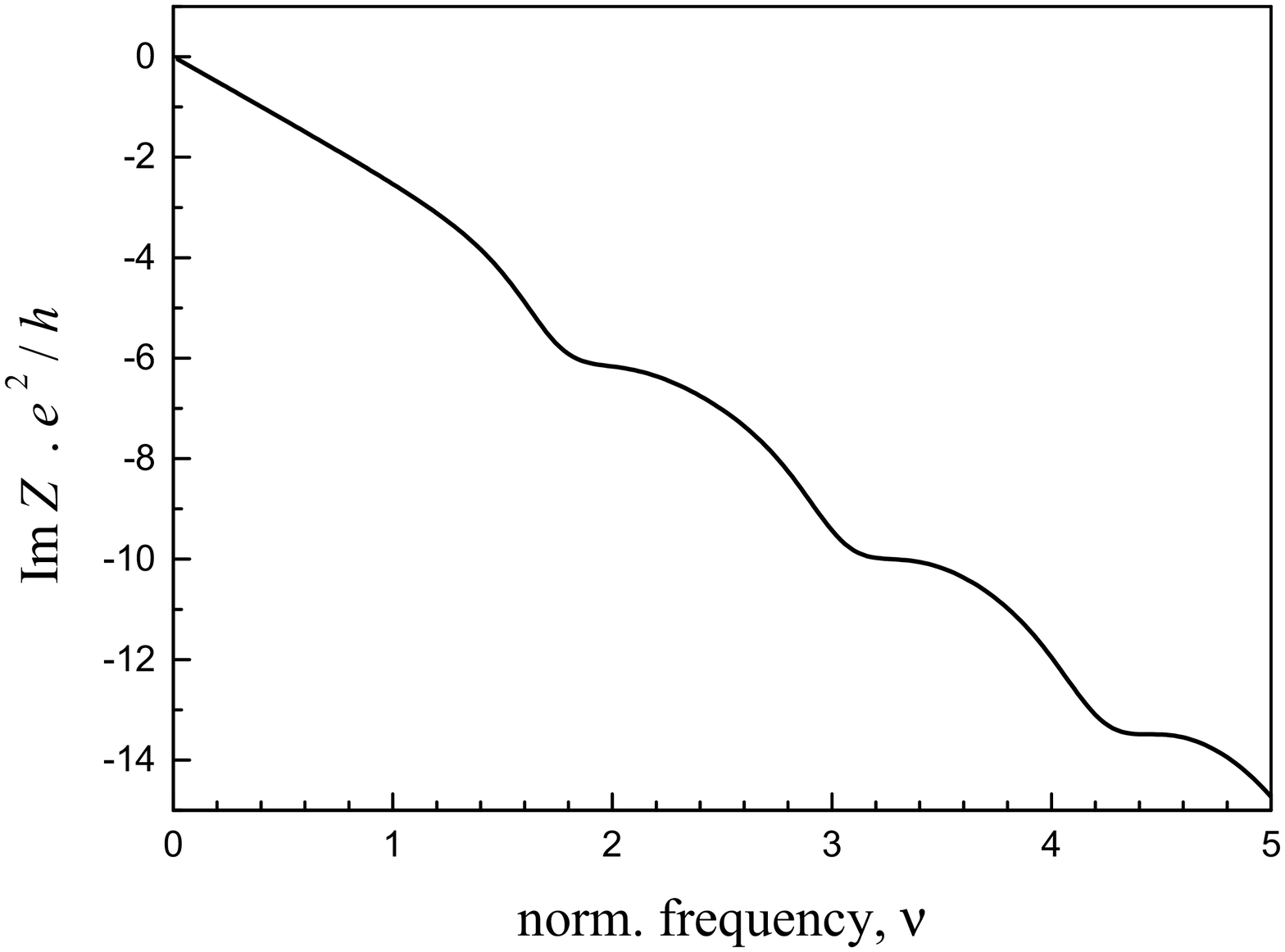,width=9.5cm}}
\caption[to]{Imaginary part of the impedance versus the normalized
frequency $\nu$.}
\label{fig6}
\end{figure}
The frequency dependence of ${\rm Im}Z$ consists in the linear decrease
caused by the first term in the brackets and oscillations around this
dependence due to the second term. This behavior is illustrated by Fig.~6.
The linear dependence of the ${\rm Im}Z$ on the frequency is obviously
dominating, which allows one to interpret $L_{\rm eff}$ as a frequency-independent
inductance.~\cite{Sablikov}

When the frequency is small, the second term on the right-hand side of
Eq.~(\ref{ImZ}) is comparable with the first one. In this case we
can expand Eq.~(\ref{ImZ}),
$$
{\rm Im}Z  = \frac{h}{e^2}\,\frac{\Omega}{2\tilde g}\:
\left[-1 + \frac{2\tilde g^2}{\pi^2} \sum_{n=1}^{\infty}
\frac{1}{n^2(1+\beta \lambda_{2n})} \right]\,,
$$
to estimate the second term. If the electron-electron interaction is omitted
($\beta=0$, $\tilde g=1$), the second term is equal to 1/3. Upon increasing
$\beta$, the second term decreases and {\rm Im}Z remains negative in spite
of the fact that the backscattering parameter $\tilde g$ slightly
increases when the interaction is turned on. Thus the reactive part of the
impedance is always inductive if the electron-electron interaction is
repulsive independent of the interaction strength.

The behavior of the impedance we have obtained here correlates with that
found for the case of both the short-range interaction~\cite{Sablikov} and
noninteracting electrons.~\cite{Sablikov3}
Somewhat different behavior of the impedance was found recently in
Ref.~\onlinecite{Cuniberti} for a quantum wire without leads based on a
rather general approach, which allows one to consider an arbitrary
distribution of the external electric field along the quantum wire.  In this
case the impedance was shown to include both the inductive and capacitive
components.  This difference originates from the different experimental
situation which was considered. In Ref.~\onlinecite{Cuniberti}, a homogeneous
wire of infinite length with a continuous spectrum of eigensolutions was
examined which results in a dispersion relation $\omega =\omega (k)$. The
resonant feature of the impedance is caused by the inflection point of
$\omega (k)$ where the group velocity reaches a minimum, with the
characteristic wavelength of the charge waves being of the order of the wire
radius. In the present paper, we consider a more specific situation of a
finite wire restricted by leads with a discrete spectrum of eigenfunctions.
The resonances we have found are attributed to the finite length of the
wire.  They appear when the characteristic wavelength of the charge waves is
of the order of the wire length, i.e., the frequency is much lower than the
resonant frequency which appears in Ref.~\onlinecite{Cuniberti}.

The total admittance of the quantum wire structure is formed by both the
quantum wire impedance $Z$ defined by Eq.~(\ref{Z}) and the interelectrode
capacitance $C_0$ which is necessarily present there. Using Eq.~(\ref{j_meas})
one obtains
\begin{equation}
\label{Y_tot}
Y_{\rm tot}(\omega)= \frac{1}{Z}-i\omega C_0 = \frac{{\rm Re}Z -
i({\rm Im}Z+\omega C_0|Z|^2)}{|Z|^2}\,.
\end{equation}

It is of interest to find the eigenfrequencies of the admittance (the
impedance) which describe the behavior of the system under consideration
as an element of electric circuit. They are known to be determined by the
poles and zeros of the admittance (the impedance). The admittance zeros
characterize the system when the external circuit is open while the poles
correspond to the short-circuit case.

Equation~(\ref{Y_tot}) shows that the poles of $Y_{\rm tot}$ coincide with the
zeros of $Z$ (or the poles of the quantum wire admittance). It follows from
Eqs.~(\ref{Z}) and (\ref{D}) that there is only one set of eigenfrequencies
which are complex with a negative imaginary part corresponding to decaying
fluctuations. Several authors have found two sets of the eigenfrequencies in
the case of short-range interacting electrons for a three-terminal
structure~\cite{Blanter} or using an other way for the calculation of the
observed current.~\cite{Ponomarenko2} Both sets of eigenfunctions also
describe the decaying fluctuations.

Of more interest, in our opinion, are the zeros of $Y_{\rm tot}(\Omega)$.
We show that in this case the conditions can be found under which the
eigenfrequencies are real and charge-wave excitations are very slowly
decaying. According to Eq.~(\ref{Y_tot}), $Y_{\rm tot} = 0$ when two equations
are satisfied simultaneously:
\begin{eqnarray}
{\rm Re}Z(\Omega) = 0 \,, \label{a}\\
{\rm Im}Z(\Omega)+\frac{\Omega v C_0}{L}\,|Z(\Omega)|^2 = 0 \label{b}\,.
\end{eqnarray}
The first equation is obviously satisfied by the set of real frequencies
$\Omega_n$ defined by Eq.~(\ref{Omega_n}). The second equation can also be
satisfied at one of frequencies $\Omega_n$ if the capacitance $C_0$ is
appropriately fitted, for instance, via changing the lead areas. The resonant
value of $C_0$ is estimated as
$$
C_0 \approx \frac{e^2 L}{2\pi^2 h v_F}.
$$
Physically, Eq.~(\ref{a}) means the absence of dissipation while
Eq.~(\ref{b}) is a resonant condition for the circuit which consists of the
quantum wire inductance and the interelectrode capacitance. Under this
condition, the resonant frequency of the charge waves in the wire coincides
with that of the $L_{\rm eff}C_0$ circuit, which results in a strong increase of
the charge-wave amplitude.

\section{Conclusion}
In this paper we have investigated the linear transport of interacting
electrons in a quantum wire of mesoscopic length with massive leads. The key
point is the full enough account of the actual Coulomb interaction inside the
wire and of three-dimensional electric field in the surrounding media. The
Coulomb interaction in a mesoscopic quantum wire includes both the
direct interaction of electrons with each other and their interaction via the
image charges induced on the leads. We have found an exact analytical
solution of the problem. This has become possible due to
({\em i}) use of the bosonization technique which is well suited to consider
the low-energy excitation of 1D interacting electrons, and
({\em ii}) solving the equation of motion for the bosonic phase field by
expansion in terms of the eigenfunctions of the electron-electron interaction
operator, which have been found for a model case where the electrodes are
plates perpendicular to the wire.

We have found that the actual Coulomb interaction affects strongly the
electron density distribution along the wire in comparison with that in the
case of short-range interacting electrons in the conventional LL model. The
nonlocal Coulomb interaction manifests itself first of all in the noticeable
increase of the charge density in the vicinity of the contacts with leads.
Here the electron density perturbation decreases exponentially with the
distance from the contact. This effect is essential when the frequency is not
too high ($\Omega < \alpha^{-1}\sqrt{2\beta}$).

Another effect of the Coulomb interaction is the renormalization of the
charge-wave velocity. Namely, in contrast with the short-range LL model
the long-range Coulomb interaction causes the frequency-dependent
renormalization of the charge-wave velocity. This effect manifests itself in
the frequency dependence of the real part of the impedance. With increasing
frequency ${\rm Re}Z$ oscillates between two limiting values which are
independent of the interaction. These are zero and $h/e^2$.  The fact that
${\rm Re}Z$ becomes zero is related to the resonances of the charge waves
along the wire length. Such resonances occur also when there is no long-range
interaction.  The Coulomb interaction causes the resonances to be
nonequidistant in frequency, which means that the charge-wave velocity is
frequency-dependent.  At low frequency the charge-wave velocity is
essentially larger than the Fermi velocity. With increasing frequency,
the charge-wave velocity decreases and tends asymptotically to the Fermi
velocity.

\section*{Acknowledgments}
We thank Markus B\"uttiker and Yaroslav Blanter for valuable discussions.
This work was supported by INTAS (Grant No. 96-0721),
the Russian Fund for Basic Research (Grant No. 96-02-18276), and the
Russian Program "Physics of Solid-State Nanostructures" (Grant No. 97-1054).

\appendix

\renewcommand {\theequation}{\thesection.\arabic{equation}}
\section{The current being measured}
\setcounter{equation}{0}
The current $j_{\rm meas}$ being detected by the measuring device in the external
circuit is equal to the charge flow in the leads. Since the currents in the
left and right leads are obviously equal each other let us consider the
current in the left electrode. It is equal to the sum of the charge flow
through the wire $j(x=-L/2)$ and the charge stored at the lead surface per
unit time:
\begin{equation}
\label{j_Q}
j_{\rm meas} =\left. j\right|_{x=-L/2} + \frac{dQ_s}{dt} \,.
\end{equation}
The charge $Q_s$ consists of two components. One is the external charge
caused by the applied voltage if the wire is absent, $Q_{\rm ext} = C_0 V_a$,
with $C_0$ being the mutual capacity of the electrodes. The other component is
the charge $Q_{\rm ind}$ induced by the charges located within the wire. The
latter is determined as
$$
Q_{\rm ind} = \int_{S} ds\, \sigma \,,
$$
where
$$
\sigma = - \frac{\epsilon}{4\pi}\frac{\partial\varphi({\bf r})}
{\partial n}\,,
$$
$\varphi $ is the potential of the charge distributed along the wire and $n$
is the outward normal to the lead surface. The potential $\varphi$ is
expressed in terms of the charge density $\rho({\bf r})$ via the Green
function $G({\bf r},{\bf r'})$ of the Laplace equation with zero boundary
conditions at the lead surfaces.

Direct calculation results in
$$
Q_{\rm ind} = -e \int_{-L/2}^{L/2} dx' \rho (x',t) \left.
\frac{\partial G_q(x,x')}{\partial x}\right|_{x=-L/2, q=0} \,.
$$
Using Eq.~(\ref{G_q(x,x')}) and the continuity
equation
$$
\frac{\partial \rho}{\partial t} = - \frac{\partial j}{\partial x}\,,
$$
we obtain the induced current
\begin{equation} \label{j_ind}
\frac{dQ_{\rm ind}}{dt} = -j(-\frac{L}{2},t) + \frac{1}{L}
\int_{-L/2}^{L/2}dx'j(x',t)\,,
\end{equation}
where $j(x,t)$ is the particle current in the wire.

Combining Eqs.~(\ref{j_Q}) with (\ref{j_ind}) one obtains
\begin{equation} \label{j_meas}
j_{\rm meas} = \frac{1}{L}\int_{-L/2}^{L/2}dx'j(x',t) + C_0 \frac{dV_a}{dt}\,.
\end{equation}
The first term in equation (\ref{j_meas}) is a current induced by electrons
moving in the wire, while the second one is a trivial capacitance current.
Equation (\ref{j_meas}) is a particular case of the general Shockley
theorem.~\cite{Shockley} For arbitrary form and configuration of the leads
the Shockley theorem is presented as follows
\begin{equation}
\label{j_meas1}
j_{\rm meas} = \frac{1}{V_a}\int_{-L/2}^{L/2}dx'j(x',t) F(x') + C_0
\frac{dV_a}{dt}\,,
\end{equation}
where $V_a$ is the potential difference between the leads and $F(x)$ is
the electric field along the electron trajectory due to $V_a$. According to
the original derivation,~\cite{Shockley} the field $F(x)$ appears here as a
result of using the reciprocity theorem when calculating the charge induced
in the leads by the charges moving along the wire. Thus, $F(x)$ has a sense
of the external electric field which does not include the polarization of the
1D electron system.

Recently~\cite{Kawabata,Oreg} a question was discussed regarding which
electric field determines the measured electric current in the quantum wire
-- the external field or the internal one -- which depends on the
polarization of 1D electrons.  In this connection we note that in our case
the current calculated according to Eq.~(\ref{j_meas1}) does not depend on
what electric field is used.  The internal electric field is defined by the
right-hand side of Eq.~(\ref{Phi(x,t)}), where the first term is the external
field $F_{ext}$ and the second one is the induced field $F_{ind}$. It is easy
to see using Eqs.~(\ref{j(x)}), (\ref{u'M}), and (\ref{uM}) that the integral
of the product $j(x) F_{ind}(x)$ is zero:
$$
\int\!\!\!\int_{-L/2}^{L/2}dx\,dx'\,j(x,t) \frac{\partial V(x,x')}{\partial
x}\, \frac{\partial \Phi(x')}{\partial x'}\,=\,0\,.  $$ Since in our case
$F_{ext}$ is independent of $x$, the measured current can be found from
Eq.~(\ref{j_meas}).


\begin{thebibliography}{30}
\bibitem[*]{e-mail} Electronic address: vas199@ire216.msk.su

\bibitem{Haldane} F.D.M.~Haldane, J. Phys. C {\bf 14}, 2585 (1981).

\bibitem{Voit} J.~Voit, Rep. Prog. Phys. {\bf 58}, 977
(1995).

\bibitem{Tarucha} S.~Tarucha, T.~Honda, and T.~Saku, Solid State
Commun. {\bf 94}, 413 (1995).

\bibitem{Maslov} D.L.~Maslov and M.~Stone, Phys. Rev. B {\bf 52},
R5539 (1995).

\bibitem{Ponomarenko1} V.V.~Ponomarenko, Phys. Rev. B {\bf 52}, R8666
(1995).

\bibitem{SafiSchulz} I.~Safi and H.J.~Schulz, Phys. Rev. B {\bf 52}, R17 040
(1995).

\bibitem{Yacoby1} A.~Yacoby, H.L.~Stormer, N.S.~Wingreen, L.N.~Pfeiffer,
K.W.~Baldwin, and K.W.~West, Phys. Rev. Lett. {\bf 77}, 4612 (1996).

\bibitem{Yacoby2} A.~Yacoby, H.L.~Stormer, K.W.~Baldwin, L.N.~Pfeiffer,
and K.W.~West, Solid State Commun. {\bf 101}, 77 (1997).

\bibitem{GlazmanRuzinShklovskii}L.I.~Glazman, I.M.~Ruzin, and B.I.~Shklovskii,
Phys. Rev. B {\bf 45}, 8454 (1992).

\bibitem{Egger} R.~Egger and H.~Grabert, Phys. Rev. Lett. {\bf 79}, 3463
(1997).

\bibitem{Blanter} Ya.M.~Blanter, F.W.J.~Hekking, and M.~B\"uttiker,
Phys. Rev. Lett. {\bf 81}, 1925 (1998).

\bibitem{Buttiker} M.~B\"uttiker and T.~Christen, in Quantum Transport in
Semiconductor Submicron Structures, edited by B. Kramer, Vol. {\bf 326} of
NATO Advanced Study Institute Series E (Kluwer, Dordrecht, 1996) pp. 263-291.

\bibitem{Sandler} N.P.~Sandler and D.L.~Maslov, Phys. Rev. B {\bf 55}, 13 808
(1997).

\bibitem{Ponomarenko2} V.V.~Ponomarenko, Phys. Rev. B {\bf 54}, 10 328
(1996).

\bibitem{Sablikov} V.A.~Sablikov and B.S.~Shchamkhalova, Pis'ma Zh. Eksp.
Teor. Fiz. {\bf 66}, 40 (1997) [JETP Lett. {\bf66}, 41 (1997)].

\bibitem{Cuniberti} G.~Cuniberti, M.~Sassetti, and B.~Kramer,
Phys. Rev. B {\bf 57}, 1515 (1998).

\bibitem{SablShch} V.A.~Sablikov and B.S.~Shchamkhalova, Pis'ma Zh. Eksp.
Teor. Fiz. {\bf 67}, 184 (1998) [JETP Lett. {\bf 67}, 196 (1998)].

\bibitem{Schulz} H.J.~Schulz, Phys. Rev. Lett. {\bf 71}, 1864 (1993).

\bibitem{SabPol1} V.A.~Sablikov and S.V.~Polyakov, unpublished.

\bibitem{SablPol} V.A.~Sablikov and S.V.~Polyakov, Phys. Low-Dimens. Struct.
{\bf 5/6}, 101 (1998).

\bibitem{DasSarma} S.~Das Sarma and E.H.~Hwang, Phys. Rev. B {\bf 54}, 1936
(1996).

\bibitem{Shockley} W.~Shockley, J. Appl. Phys. {\bf 2}, 635 (1938).

\bibitem{Velicky} B.~Velicky, J.~Masek, and B.~Kramer, Phys. Lett. A {\bf
140}, 447 (1989).

\bibitem{Sablikov3} V.A.~Sablikov and E.V.~Chensky, Pis'ma Zh. Eksp.
Teor. Fiz. {\bf 60}, 397 (1994) [JETP Lett. {\bf 60}, 410 (1994)].

\bibitem{Kawabata} A.~Kawabata, J. Phys. Soc. Jpn. {\bf 60}, 30 (1996).

\bibitem{Oreg} Y.~Oreg and A.M.Finkel'stein, Phys. Rev. B {\bf 54}, R14 269
(1996).

\end{thebibliography}
\end{document}